\documentclass[%
 reprint, prl,
superscriptaddress,
 amsmath,amssymb,
 aps,
]{revtex4-1}

\usepackage{color}
\usepackage{graphicx}
\usepackage{dcolumn}
\usepackage{bm}
\usepackage{fdsymbol}
\usepackage{bbold}

\newcommand{\beginsupplement}{%
        \setcounter{table}{0}
        \renewcommand{\thetable}{S\arabic{table}}%
        \setcounter{figure}{0}
        \renewcommand{\thefigure}{S\arabic{figure}}%
        \setcounter{equation}{0}
        \def\theequation{S\arabic{equation}}
     }

\newcommand{\rev}[1]{#1}

\renewcommand{\vec}{\boldsymbol}

\begin{document}

\preprint{APS/123-QED}

\title{Optimal noise-canceling networks}

\author{Henrik Ronellenfitsch}
 \email{henrikr@mit.edu}
\author{J\"orn Dunkel}
\email{dunkel@mit.edu}
\affiliation{%
Department of Mathematics, Massachusetts Institute of Technology, Cambridge, MA, U.S.A.
}%

\author{Michael Wilczek}
\email{michael.wilczek@ds.mpg.de}
\affiliation{%
Max Planck Institute for Dynamics and Self-Organization, G\"ottingen, Germany
}%

\date{\today}

\begin{abstract}
Natural and artificial networks, from the cerebral cortex to large-scale  power grids,  face the challenge of converting noisy inputs into robust signals. The input fluctuations often exhibit complex yet statistically reproducible correlations that reflect underlying internal or environmental processes such as synaptic noise or atmospheric turbulence. This raises the practically and biophysically relevant question of whether and how noise-filtering can be hard-wired directly into a network's  architecture. By considering generic phase oscillator arrays under cost constraints, we explore here analytically and numerically the design, efficiency and topology of noise-canceling networks. Specifically, we find that when the input fluctuations become more correlated in space or time, optimal network architectures become sparser and more hierarchically organized, resembling the vasculature in plants or animals. More broadly, our results provide concrete guiding principles for designing more robust and efficient power grids and sensor networks.

\end{abstract}

\pacs{Valid PACS appear here}
\maketitle


Fluctuations fundamentally limit the function and efficiency of physical~\cite{LINDNER2004321} and biological~\cite{Bialek1987,Tsimring2014} networks across a wide spectrum of scales. Important examples range from atmospheric turbulence~\cite{Wyngaard2010,Milan2013}
affecting large telescope arrays~\cite{Wootten:2009aa}, wind farms~\cite{Sorensen2002,Luo2007,Lukassen2018,Chen2005,Tande2003} and power grids~\cite{Schmietendorf2017,Schafer2018,Nesti2018,Haehne2018,Coletta2018} to
neuronal noise in the auditory~\cite{DERIDDER201416,LEAVER201133} and visual~\cite{Averbeck2006,Kanitscheider2015} cortices,
and extrinsic and intrinsic fluctuations~\cite{hilfinger2011separating} in gene expression pathways~\cite{thattai2001intrinsic,paulsson2004summing}.
Over the last decades, remarkable progress has been made in the development  and understanding of noise-suppression strategies~\cite{chua1988cellular,moon1993coordinated}, and their limits~\cite{Bialek1987,lestas2010fundamental} in physical~\cite{Tande2003,Li2017,Tyloo2018} and biological~\cite{DERIDDER201416,Averbeck2006,Zechner2016} networks.
\rev{Classical adaptive noise filtering~\cite{Bucy1967,Widrow1975,Clarkson1993} utilizes  active  control~\cite{Klickstein2017,Yan2015},
and networks can be optimized for active  controllability~\cite{Xiao2014,Liang2016,Wang2012}
and/or transport efficiency~\cite{Banavar2000,Katifori2010,Bohn2007,Segre2002,Durand2006,Andrews2006}.
}
Still lacking at present are generic design principles for the construction of optimal \rev{passive} noise-canceling networks (NCNs). \rev{While
passive noise-reduction has been demonstrated for single oscillators~\cite{Kenig2012}}, it is not yet well understood how the architecture and efficiency of optimal NCNs depends on the input correlations and constraints in natural and man-made systems. Deciphering these dependencies can yield
more robust sensory network and power grid designs and may also help clarify the role of noise-reduction in biological network evolution.

Correlated input fluctuations can have profound biomedical or technological consequences in hierarchical network structures. For instance, the detection neurons of the retina are subject to correlated fluctuations~\cite{Ala-Laurila2011} which are passed on to the visual cortex where input noise has been
shown to affect neural processing~\cite{Averbeck2006}. Similarly, deficient noise-cancellation in dysfunctional auditory sub-networks has been proposed as a potential cause of tinnitus~\cite{DERIDDER201416,LEAVER201133}.
Another conceptually related problem of rapidly increasing importance is the feed-in of
spatio-temporally correlated power fluctuations from solar and wind farms into multi-national power grids~\cite{Nesti2018,Schafer2017,Schafer2018,Schmietendorf2014,Schmietendorf2017,Tande2003,Manik2017,Sorensen2002,bandi2017spectrum,Milan2013,Lukassen2018}. These examples raise the general question to which extent  efficient noise-cancellation \rev{can be hard-wired into a network's architecture if the signal fluctuations have known statistics}.

\par
Here, we show both analytically and numerically for generic oscillator networks~~\cite{Schmietendorf2017,Nardelli2014,Schafer2017,Acebron2005,Kromer2017} that it is indeed possible to design optimized weighted network
topologies capable of suppressing `colored' fluctuations~\cite{hanggi1994colored,Lukassen2018} as typically present in biological and engineered systems. In stark contrast to the widely studied problem of optimal synchronization~\cite{Rodrigues2016,Bag2007,Gray1994,Tanaka2008,Fazlyab2017,Skardal2014,Fardad2014,Kelly2011,Li2017,Brede2008,Meng2018}, our results imply that optimal NCNs harness desynchronization to reduce  fluctuations globally.  Importantly, NCNs operate purely passively, canceling out
a substantial fraction of the input fluctuations without requiring active
smoothing---the network itself acts as the filter. As a general principle,
we find that the more correlated fluctuating inputs are in space or time, the sparser
and the more hierarchically organized the NCN will be. Interestingly, the best-performing networks are often
reminiscent of  leaf venation or animal vasculature, supporting the view that robustness against fluctuations has been an  evolutionary factor~\cite{Katifori2010,Hu2013}.  The mathematical analysis below thus provides detailed guidance for how to use bio-mimetic network topologies to improve noise-robustness in engineered grids and sensor networks.

\begin{figure*}[t]
\includegraphics[width=\textwidth]{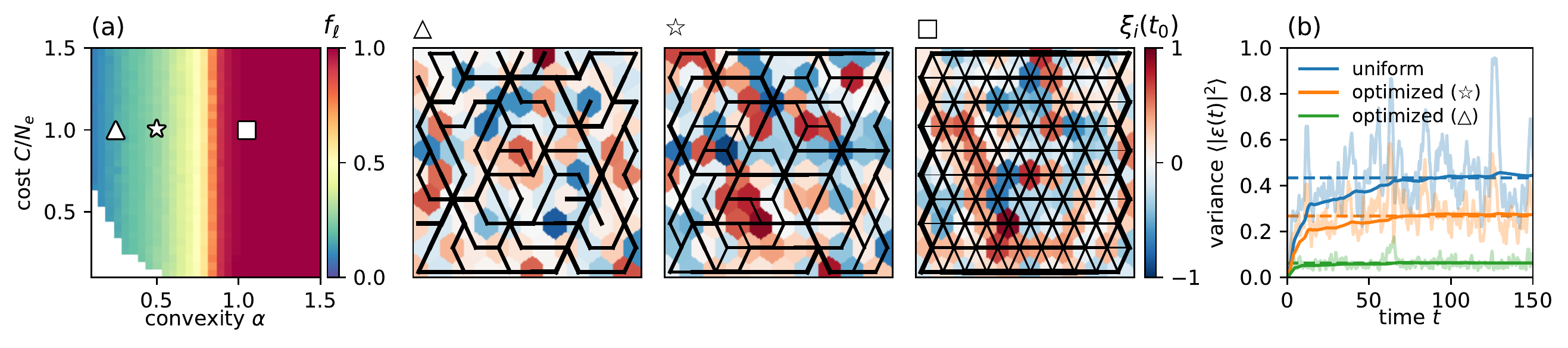}
\caption{Even for spatially incoherent white noise input $\sigma,\tau\to 0$, optimal NCNs  exhibit a non-trivial sparse topology independent of the nonlinear steady state.
(a) The fraction of loops $f_\ell = N_\ell/N_\mathrm{tri}$, where $N_\ell$ is the number of loops in the optimized network and
$N_\mathrm{tri}$ is the number of loops in a triangular grid, measures
the topology of optimal networks. Each of the $30\times 30$ pixels in the cost-convexity phase diagram is an average over 15 optimal
networks obtained for different uniformly random initial $B_{ij}$. In the white domain, no solutions to Eq.~\eqref{eq:nonlinear-fp} were found.
The NCN topology $f_\ell$ is effectively independent of $C$.
Panels  $\medtriangleup$, $\medwhitestar$, $\Box$ show examples of optimal NCNs with different sparsities, with edge thicknesses
proportional to $B_{ij}^\alpha$. Backgrounds show one instance
of the spatial feed-ins $\xi_i(t_0)$ normalized to $(-1,1)$.
(b)~Time-averaged
variance $\langle{|\vec\varepsilon(t)|^2}\rangle$ and instantaneous variances $|\vec\varepsilon(t)|^2$ (faint) obtained
from numerical solutions of Eq.~\eqref{eq:swing}
on uniform and optimized network topologies for
$\alpha=0.25$ ($\medtriangleup$) and $\alpha=0.5$ ($\medwhitestar$) with edge
cost $C=1$ \rev{and centered inputs}. Analytically  predicted
variances (dashed) agree with the simulations.
\label{fg:figure-1}}
\end{figure*}

To investigate noise-cancellation in a broadly applicable setting, we consider a generic model of spatially distributed,
nonlinearly coupled second-order phase oscillators, with phase angles $\delta_i(t)$ at each network node $i$, governed by
\begin{align}
	\ddot \delta_i = -\gamma \dot \delta_i + \sum_{j=1}^N B_{ij} \sin(\delta_i-\delta_j) + P_i(t),
    \label{eq:swing}
\end{align}
where $\gamma$ is a damping coefficient.  The oscillator couplings are symmetric, $B_{ij} = B_{ji}$,
and $P_i(t)$ is the fluctuating net signal or power input at site~$i$. Equation~\eqref{eq:swing}
has been successfully applied to describe the dynamics of power grids~\cite{Nardelli2014}. The Kuramoto model~\cite{Acebron2005,Rodrigues2016} is recovered in the overdamped limit, for which all subsequently derived results remain valid after a transformation of parameters (Supplemental Information).
The fluctuating inputs can be decomposed
as $P_i(t) = \bar{P}_i + \xi_i(t)$, where
$\xi_i(t)$ are the fluctuations around the constant mean $\bar P_i$.
\rev{Because Eq.~\eqref{eq:swing} is invariant under a constant shift $\delta_i\rightarrow \delta_i + c$, it is possible to split off the irrelevant dynamics of the mean $\frac{1}{N}\sum_j \delta_j$
(Supplemental Information).
As a result, only the centered inputs $\bar P_i^c = \bar P_i - \frac{1}{N}\sum_j\bar P_j$ and $\xi_i^c = \xi_i - \frac{1}{N}\sum_j \xi_j$
are relevant.}
\rev{Adopting this mean-centered frame of reference from now on}, we write
$\delta_i(t) = \bar \delta_i + \varepsilon_i(t)$ for constant average
phase angles $\bar\delta_i$ and fluctuations $\varepsilon_i(t)$.
Assuming that the angle fluctuations $\varepsilon_i(t)$
are small and linearizing around
$\bar \delta_i$, we obtain the coupled set of equations,
\begin{align}
	0 &= \sum_{j=1}^N B_{ij} \sin(\bar \delta_i - \bar \delta_j) + \rev{\bar P_i^c} \label{eq:nonlinear-fp} \\
    \ddot\varepsilon_i  &= -\gamma \dot\varepsilon_i + \sum_{j=1}^N \left[B_{ij} \cos(\bar\delta_i - \bar\delta_j)\right] (\varepsilon_i - \varepsilon_j) + \rev{\xi_i^c(t)}.
    \label{eq:linearized}
\end{align}
The zeros of the nonlinear algebraic Eq.~\eqref{eq:nonlinear-fp}
correspond to fixed points of the dynamics Eq.~\eqref{eq:swing}.
Our main goal here is to use Eq.~\eqref{eq:linearized} to derive and characterize
optimal couplings $B_{ij}$ that \rev{minimize the total fluctuation variance $\langle | \vec\varepsilon(t) |^2 \rangle$, where the vector $\vec\varepsilon(t)$ has
components $\varepsilon_i(t)$, the total instantaneous
variance is the norm $| \vec\varepsilon(t) |^2$, and $\langle \,\cdot\,\rangle$ denotes a time average.}
The optimal network connectivity $B_{ij}$  will depend on the statistics of the input fluctuations, encoded in the elements $R_{ij}(t,t') = \langle \xi_i(t) \xi_j(t) \rangle$ of the covariance matrix~$R$.
\par

Throughout, we assume that spatio-temporal correlations
factorize, although the general approach extends to the non-factorizing case. For the time-correlations,
we focus on colored Ornstein-Uhlenbeck noise~\cite{hanggi1994colored}
with $R(t,t')= \hat R\, e^{-|t-t'|/\tau}/(2\tau)$.
In the limit of correlation time $\tau\rightarrow 0$,
white noise is recovered with $R(t,t') = \hat R\, \delta(t-t')$. For the spatial part $\hat R=(\hat R_{ij})$, we choose generic
isotropic and homogeneous Gaussian covariances
$G_{ij} = e^{-|\mathbf{x}_i - \mathbf{x}_j|^2/(2\sigma^2)}$,
where $\mathbf x_i$ is the spatial position of oscillator $i$ and
$\sigma$ is a correlation length.
In the limit $\sigma\rightarrow 0$,
the feed-ins become incoherent with \rev{$\hat R_{ij} = \delta_{ij}$.}
The total fluctuation variance $\langle | \vec{\varepsilon}(t) |^2 \rangle$ can be
calculated analytically for any $\hat R$ in the Langevin formalism
(Supplemental Information),
\begin{align}
	\langle | \vec\varepsilon(t) |^2 \rangle =
    \frac{1}{2\gamma} \operatorname{tr}\left(
     \left[\mathbb{1} + \frac{\tau^2}{1+\gamma\tau}L\right]^{-1}  L^\dagger  \hat R \right),
    \label{eq:objective-ou}
\end{align}
where $L$ is the weighted graph Laplacian matrix
of the network with the weights of edge $(ij)$ given by $B_{ij}\cos(\bar\delta_i-\bar\delta_j)$,
\rev{and $\operatorname{tr}(\cdot)$ is the matrix trace.
The pseudo-inverse $L^\dagger$ intrinsically acts as a projection to center $\hat R$.}
In the white-noise limit $\tau\rightarrow 0$, Eq.~\eqref{eq:objective-ou} reduces to
\begin{align}
	\langle | \vec\varepsilon(t) |^2 \rangle = \frac{1}{2\gamma}
\operatorname{tr}\left( L^\dagger \hat R \right).
\label{eq:objective-wn}
\end{align}
\begin{figure*}
\includegraphics[width=.8\textwidth]{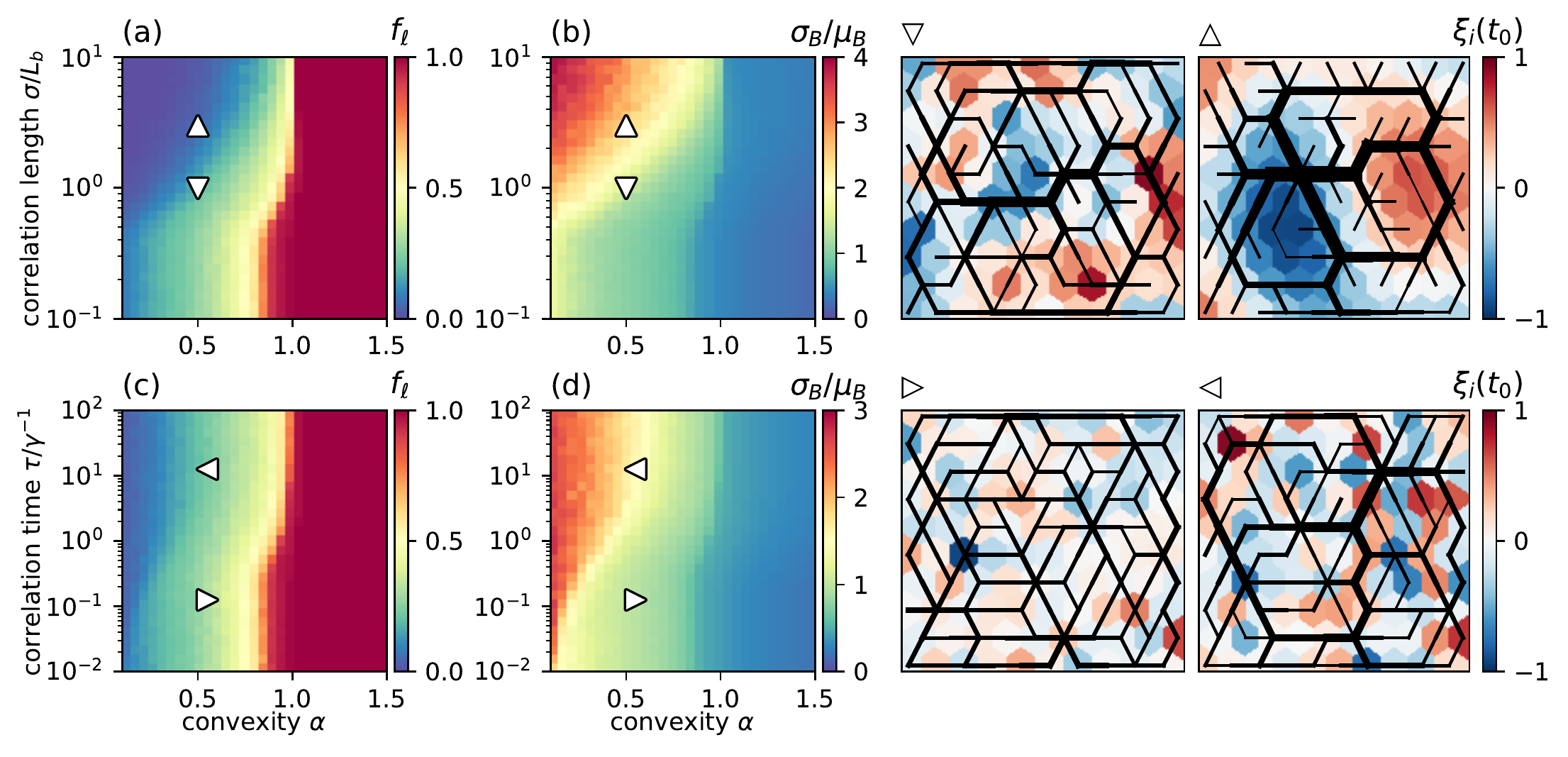}
\caption{Spatial and temporal input correlations lead to a similar hierarchical NCN organization
despite acting through different mechanisms.
(a,b)~Gaussian spatial correlations $\sigma>0$ with temporal white noise $\tau\to 0$.
The loop fractions $f_\ell$ in (a)  show  that NCN topology depends largely on
$\alpha$, although  the transition between
loopy and sparse phase shifts when the correlation scale~$\sigma$ approaches the mean edge length $L_b$.
For $\sigma\gg L_b$ networks become sparser when $\alpha\sim 1$.
(b)~The coupling variance $\sigma_B$, normalized by the mean $\mu_B$, indicates  that non-uniform hierarchical patterns and sparsity are strongly correlated.
(c,d)~Ornstein-Uhlenbeck colored noise $\tau>0$ with spatially incoherent feed-ins $\sigma \to 0$
shows hierarchical patterns similar to those in panels~(a,b).
Examples of optimal networks  at the  positions marked by symbols in the phase diagrams illustrate the transitions from  dense uniform networks to sparse hierarchical networks with increasing spatial or temporal correlation.
Each of the $30\times 30$ pixels in (a--d) is an average over 15 optimal networks.
\label{fg:figure-2}}
\end{figure*}
The structure of Eqs.~\eqref{eq:objective-ou} and~\eqref{eq:objective-wn} implies
that, in principle,  arbitrarily small variances
$\langle | \vec\varepsilon(t) |^2 \rangle$
can be achieved by choosing the $B_{ij}$ arbitrarily large.
In natural or engineered real-word networks, however,  the allowed values of the $B_{ij}$ are
 restricted by construction or maintenance costs. To account for this fact, we adopt here the widely used~\cite{Tanaka2008,Katifori2010,Bohn2007,Hu2013,Ronellenfitsch2016}  cost constraint
$\sum_{(ij)} B_{ij}^\alpha = N_e C$,  where $\alpha>0$ is a convexity parameter, $C$ the cost per edge, and $N_e$  the number of edges in the network.
In the concave regime $\alpha <1$, one expects sparse networks because it becomes more economical
to construct a single edge with a large coupling rather than to distribute
over, say, two smaller ones. Since many natural networks are sparse, and
sparsity is desirable in engineering, this concave range arguably comprises the
most interesting part of phase space.
The cost-constrained optimization
\rev{is carried out starting from a given base network and
initial $B_{ij}$.
Optimal weights are found iteratively based on the method of
Lagrange multipliers (Supplemental Information).
Weights $B_{ij}=0$ in the final optimized network
correspond to edges being pruned from the base network, and
thus to changes in topology.}
In the case of white noise in time and close to synchrony ($\bar \delta_i
\approx 0$), the minima have
an interesting interpretation: using the eigen-decomposition
$\hat R = \sum_k \rho_k \vec r_k \vec r_k^\top$, one finds the defining relation
$\alpha \lambda B_{ij}^{\alpha+1} = \sum_k \rho_k \, [B_{ij} (\varepsilon_i^{(k)} -
\varepsilon_j^{(k)})]^2$,
where the
$\vec\varepsilon^{(k)}$ are steady-state angles in the presence
of steady feed-ins~$\vec r_k$. Thus, the optimal
couplings are directly related to a weighted average
over \emph{local} steady state flows.
In the general case, additional terms appear (Supplemental Information).
Armed with these analytical insights, we now turn to the numerical investigation
of optimal NCNs for different input noise statistics.

We explore planar triangular grids \rev{as base networks}
as approximately realized in many biological and engineering systems
such as cilia~\cite{Hudspeth2008,drescher2010fidelity} or staggered wind farms~\cite{Stevens2017}.  The number of nodes is $N = 100$ and damping fixed at $\gamma=0.5$, following Ref.~\cite{Nardelli2014}. \rev{The uncentered steady feed-ins are
$\bar P_i = \eta_i$, where the $\eta_i$
are independent Gaussian random variables with zero mean and unit variance.
}
Covariance matrices are normalized to $\operatorname{tr}(\hat R) = 1$,
bringing steady state background and fluctuations to a similar scale.
Numerical solutions of Eq.~\eqref{eq:swing}
were obtained using the Euler-Maruyama scheme with time
step $\Delta t = 10^{-3}$. All main results remain valid for other grid geometries as well~(Supplemental Information).

\begin{figure*}[th]
\includegraphics[width=\textwidth]{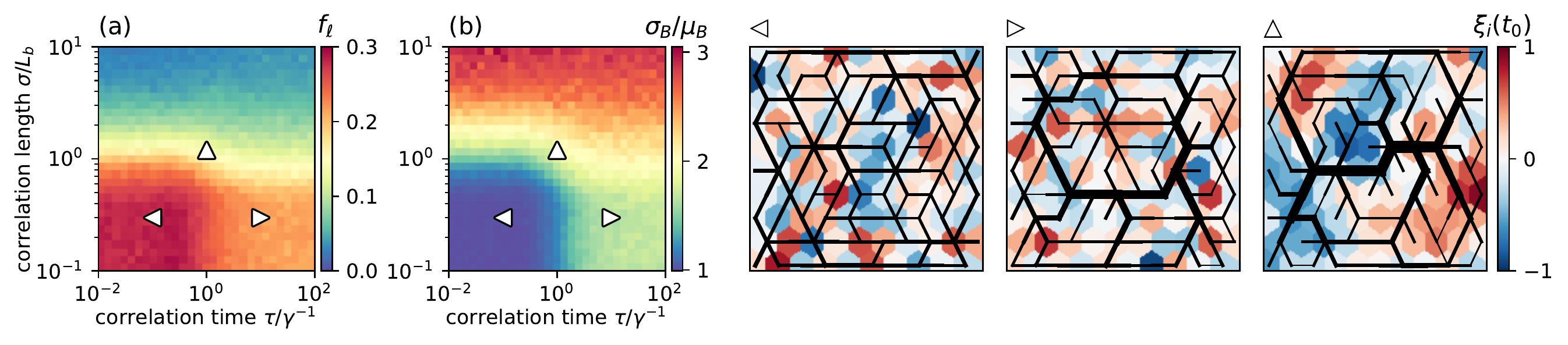}
\caption{Combining spatial and temporal correlations leads to three qualitatively
distinct NCN phases in the $(\tau,\sigma)$-plane.
(a)~The loop density $f_\ell$ characterizes the three phases as follows:
Short correlation times $\tau$  and short correlation lengths $\sigma$ favor highly reticulate redundant networks  ($\medtriangleleft$),
large $\tau$ and small $\sigma$ lead to a moderate reticulation ($\medtriangleright$), whereas large $\tau$ and  large $\sigma$ selects low reticulation ($\medtriangleup$).
(b)~The coupling spread $\sigma_B/\mu_B$ indicates a similar division of the $(\tau,\sigma)$-phase plane:
Low $\tau, \sigma$ lead to highly uniform networks ($\medtriangleleft$),
high $\tau$ and low $\sigma$ lead to networks with an intermediate coupling variability ($\medtriangleright$), and
high $\tau, \sigma$ lead to strongly hierarchical networks with large spread in the couplings $B_{ij}$ ($\medtriangleup$).
The three phases are separated approximately by the lines $\tau/\gamma^{-1} \sim 1$ and $\sigma/L_b \sim 1$.
Each pixel in the $30\times 30$ plots (a,b) is an average over 15 optimal networks; $\alpha=0.5$, $C=1$ in all panels.
\label{fg:figure-3}}
\end{figure*}

Already in the simplest case, when node inputs are white noise in time ($\tau\to 0$) and spatially incoherent
($\sigma\to 0$), optimal NCNs exhibit non-trivial topologies
in the sparse regime $0<\alpha<1$ [Fig.~\ref{fg:figure-1}].
The fraction of loops $f_\ell = N_\ell/N_{\mathrm{tri}}$,
where $N_\ell$ is the number of loops in the optimal network
and $N_{\mathrm{tri}}$ is the number of loops in the full
triangular grid, decreases with $\alpha$  [Fig.~\ref{fg:figure-1}(a)]. This indicates that optimal NCNs become sparser for $\alpha\to 0$. The \rev{nonzero couplings in the optimized network have similar magnitude} for uncorrelated inputs [Fig.~\ref{fg:figure-1}(a,$\medtriangleup$,$\medwhitestar$)],
\rev{and the optimal networks do not follow any symmetry of the base network}. As expected, optimal networks become dense for $\alpha>1$~[Fig.~\ref{fg:figure-1}(a, $\square$)] \rev{and
retain the base network topology}.  The nonlinear steady state, despite being fully taken into
account in our optimization procedure, has little influence on the structure of optimal NCNs.
Decreasing the mean coupling through the cost $C$ pushes the NCNs
\rev{towards the regime lacking solutions of Eq.~\eqref{eq:nonlinear-fp}} but causes no significant
changes in topology apart from an overall scaling
of the couplings, even very close to the transition [Fig.~\ref{fg:figure-1}(a)].
Simulations of the full nonlinear Eq.~\eqref{eq:swing} on the identified sparse
NCNs confirm a significant noise reduction compared to \rev{uniform weights}, in quantitative agreement with the predictions  of the linear model [dashed and solid lines in Fig.~\ref{fg:figure-1}(b)]. In general, the linear approximation
is accurate as long as the noise is small compared to a worst-case uniform distribution $\varepsilon_i \in [-\pi,\pi]$
(Supplemental Information).
Since the optimal topologies show little dependence on the nonlinear
steady state [Fig.~\ref{fg:figure-1}(a)], it suffices to focus on the synchronized limit $\bar\delta_i =0$
and $C=1$  when considering correlated noisy inputs in the remainder.
The existence of non-trivial optimal NCN topologies even for uncorrelated inputs
is remarkable, and may already have practical applications.

\par
Even more interesting hierarchical NCN structures arise when the input noise becomes correlated [Fig.~\ref{fg:figure-2}]. The optimal  couplings settle into non-uniform patterns containing loopy backbone structures with tree-like branches, reminiscent of plant~\cite{Sack2013,Ronellenfitsch2015b}, fungal~\cite{Heaton2012b}, or animal~\cite{Fruttiger2002} vasculature [Fig.~\ref{fg:figure-2}($\medtriangleup$)]. To dissect the effects of correlations, we first consider fluctuating inputs that are still uncorrelated in time ($\tau\to 0$) but have a finite correlation length $\sigma>0$. Our numerical analysis shows that the topology of optimal NCNs changes as $\sigma$ is varied relative to the mean edge length $L_b$, the latter defining the natural resolution scale for a network. As expected, for $\sigma\ll L_b$, we find the same NCN topology as for incoherent inputs [cf. Fig.~\ref{fg:figure-1} and~\ref{fg:figure-2}].  In contrast, when $\sigma$  becomes comparable to or larger than $L_b$, the optimal NCNs become significantly sparser for $0 < \alpha < 1$  [Fig.~\ref{fg:figure-2}(a)]. This transition is accompanied by the gradual emergence of a hierarchical  network structure,  reflected by an increased standard deviation $\sigma_B$ of the optimal coupling parameters $B_{ij}$ relative to their mean $\mu_B$ [Fig.~\ref{fg:figure-2}(b)]. Thus, NCNs for spatially correlated white noise develop hierarchical sparse architectures as the correlation length $\sigma$ increases.
\par

These observations can be rationalized by noting that in the limit of large $\sigma$,  we have \rev{$\hat R \sim D$
where $D$ is the matrix of squared Euclidean distances between oscillators.}
The rank of $D$ is at most the dimension $d$ of the embedding space~\cite{Dokmanic2015}. Therefore, the objective Eq.~\eqref{eq:objective-wn} becomes equivalent to an average over at most $d$ steady-state inputs.
For networks with a single \rev{non-fluctuating} input, it is known that the optimal topology is a maximally sparse
tree~\cite{Banavar2000}. Since $d=2$ in our case, the optimal NCNs are close to such trees.
This argument holds for any sufficiently well-behaved $\hat R =
f(D/\sigma^2)$ that depends on the node distances via a scale parameter.
The emergence of the hierarchical structure follows from the earlier stated fact that
couplings become proportional to a mean flow, which in a tree-like
topology of steady inputs accumulates as the network graph
is traversed upstream from a leaf node.
Remarkably, for large $\sigma$, the optimal NCNs often exhibit
spontaneous  symmetry-breaking by approximately realizing \emph{rooted}
trees, in which a hierarchical backbone emanates from one or two central
nodes [Fig.~\ref{fg:figure-2}($\medtriangleup$)] even though
no such distinguished node(s) were initially prescribed.

Interestingly, colored noise with non-vanishing correlation time
$\tau>0$  but no spatial coherence ($\sigma\to 0$) has qualitatively
similar effects on the network structure.
When $\tau$ is larger than the
damping timescale $\gamma^{-1}$, optimal NCNs also become
sparser and more hierarchically patterned [Fig.~\ref{fg:figure-2}(c,d) and ($\medtriangleleft$, $\medtriangleright$)].
The origin of sparsity is now different because
$\hat R$ is almost full rank for $\sigma\to 0$, and related to
the large-$\tau$ asymptotic behavior of the objective, $\langle |\vec{\varepsilon}(t)|^2
\rangle \sim \operatorname{tr}((L^\dagger)^2 \hat R)/(2\tau)$.
Although the objective
does not scale homogeneously with $C$ anymore, only the transition between the different NCN topologies changes (Supplemental Material).


Lastly, combining spatial and temporal correlations, the $(\tau,\sigma)$-plane subdivides into three distinct phases
[Fig.~\ref{fg:figure-3}(a,b)].  For $\sigma\ll L_b$  and $\tau\ll \gamma^{-1}$,
 optimal NCNs are highly dense and uniform [Fig.~\ref{fg:figure-3}($\medtriangleleft$)].
For $\sigma\ll L_b$ but  $\tau\gg \gamma^{-1}$, NCNs exhibit intermediate sparsity and hierarchical patterning [Fig.~\ref{fg:figure-3}($\medtriangleright$)]. For $\sigma\gg L_b$,  NCNs become generally sparse and
hierarchically patterned  with little dependence on $\tau$
[Fig.~\ref{fg:figure-3}($\medtriangleup$)],  although the transition between the different NCNs topologies is shifted to smaller $\sigma$ when $\tau \gg \gamma^{-1}$.

To conclude, the above analytical and numerical results show that
noise-cancellation can be hard-wired into weighted network topology for both
uncorrelated and correlated input fluctuations. As a general rule, the more correlated the
input fluctuations, the sparser and more hierarchically ordered the optimal networks become.
Previous work~\cite{LINDNER2004321,Acebron2005} has demonstrated the applicability of the underlying phase oscillator framework to a myriad of physical and biological systems, from neuronal networks~\cite{Gray1994,Penn2016} and ciliary carpets~\cite{Niedermayer2008,uchida2010synchronization,brumley2012hydrodynamic}  to renewable energy farms and power grids~\cite{Nardelli2014,Schmietendorf2017,Lukassen2018,Nesti2018}. One can therefore expect that the above ideas and results have conceptual and practical implications for most, if not all, of these systems.

\begin{acknowledgments}
This work was supported by an Edmund F. Kelly Research Award (J.D.) and a James S. McDonnell Foundation Complex
Systems Scholar Award (J.D.).
\end{acknowledgments}

\bibliography{references}

\beginsupplement
\onecolumngrid
\section{Supplemental Material}
\section{Centered dynamics}
Eq. (1) from the main paper contains a freedom
of re-defining $\delta_i \rightarrow \delta_i +c$ for some constant
$c$ corresponding to a reference angle.
Here, we fix this freedom by introducing the new variables
\begin{align*}
    \psi_i(t) &= \delta_i(t) - \mu(t) \\
    \mu(t) &=  \frac{1}{N}\sum_j \delta_j(t).
\end{align*}
Taking derivatives and plugging them into Eq.~(1), we find that they satisfy
\begin{align}
    \ddot \psi &= -\gamma \dot\psi + \sum_{j} B_{ij}
        \sin(\psi_i - \psi_j) + P_i(t) - \frac{1}{N}\sum_j P_j(t) \label{eq:fixed-dyn}\\
        \ddot \mu &= -\gamma \dot\mu + \frac{1}{N}\sum_j P_j(t), \label{eq:mean-dyn}
\end{align}
where we used $\sum_{i,j}B_{ij} \sin(\psi_i-\psi_j) = 0$ due to
antisymmetry.
Equation~\eqref{eq:fixed-dyn} is equivalent to Eq.~(1) but with
centered inputs, and Eq.~\eqref{eq:mean-dyn} describes a
stochastically forced particle with damping.
We decompose the inputs into constant means and stochastic
fluctuations, $P_i(t) = \bar P_i + \xi_i(t)$.
Without fluctuations, a steady state is only possible if
the constant forcing in Equation~\eqref{eq:mean-dyn} vanishes,
$\sum_j \bar P_j = 0$. We shall assume this to be true from
here on and focus on Eq.~\eqref{eq:fixed-dyn}, because the dynamics of the mean is
independent of the weighted network topology encoded in the $B_{ij}$.
We find the centered dynamics
\begin{align}
    \ddot \psi &= -\gamma \dot\psi + \sum_{j} B_{ij}
        \sin(\psi_i - \psi_j) + \bar P_i + \xi_i(t) - \frac{1}{N}\sum_j \xi_j(t). \label{eq:fixed-dyn-dec}
\end{align}
Equation~\eqref{eq:fixed-dyn-dec} is again simply Eq.~(1) but
with \emph{centered} fluctuations. It is only these centered fluctuations
that are relevant for optimal NCNs.
In vector form they can be written using the projection matrix $Q$ as
\begin{align*}
  Q \vec{\xi} = \left(\mathbb{1} - \frac{1}{N}J\right)\vec{\xi},
\end{align*}
where $J_{ij}=1$. Similarly, the centered correlation matrix
is
\begin{align*}
  R_c = Q \langle \vec{\xi} \vec\xi^\top \rangle Q = QRQ.
\end{align*}

\section{Derivation of the objective function}
In this section we derive the objective function for white noise and
colored noise. Note that unlike in the main paper, for notational
ease we use the inverse correlation time scale $\kappa = \tau^{-1}$.
We first consider the case of pure white noise, and then
generalize to Ornstein-Uhlenbeck colored noise.

\subsection{White Noise}
Here, we compute the variance of fluctuations directly in the
Langevin formalism.

We consider the linearized second-order system in the centered frame
from the preceding section,
\begin{align*}
  \ddot{\vec{\varepsilon}} + \gamma \dot{\vec{\varepsilon}} + L\vec\varepsilon = Q\vec{\xi}(t),
\end{align*}

where $\langle \vec \xi \rangle = \vec0$, $\langle \vec \xi(t) \vec \xi(t')^\top \rangle = QRQ\, \delta(t - t')$
is white noise input in time with spatial correlation matrix $R$.
We can rewrite the system as first order,
\begin{align*}
  \begin{pmatrix}
    \dot{\vec\varepsilon} \\
    \dot{\vec\nu}
  \end{pmatrix}
  &=
  \begin{pmatrix}
    0 & \mathbb{1} \\
    -L & -\gamma \mathbb{1}
  \end{pmatrix}
  \begin{pmatrix}
    \vec\varepsilon \\
    \vec\nu
  \end{pmatrix}
  +
  \begin{pmatrix}
    0 \\
    Q\vec \xi
  \end{pmatrix} \\
  \Leftrightarrow \dot{\vec y} &= M \vec{y} + \vec{u}.
\end{align*}
The solution to this system can be expressed as
\begin{align*}
  \vec y(t) = \exp(M t) \vec y_0 + \int_0^T \exp(M(t- t')) \vec u(t') \,dt'.
\end{align*}
The eigenvalues of $M$ are easy to compute by explicitly writing down
the eigenvector condition in block-matrix form. One obtains
\begin{align*}
  \lambda_{i,\pm} = -\frac{\gamma}{2} \pm \sqrt{\frac{\gamma^2}{4} - \omega_i^2},
\end{align*}
where the $\omega_i^2$ are the (positive) eigenvalues of the Laplacian $L$.
Since $\operatorname{Re}(\lambda_{i,\pm}) < 0$ except for the eigenvector of all
1's in the first block, the homogeneous solution
$\exp(M t) \vec y_0$ decays for large times except for a constant
angular shift. In the following, we change into a frame where this
shift vanishes and focus on the particular solution.

We want to compute the matrix of correlations for large times,
\begin{align}
  \langle \vec y(t) \vec y(t')^\top \rangle &= \int_0^T d{s} \int_0^{t'} d{s}' \exp(M(t- {s})) \langle \vec u({s}) \vec u^\top({s}') \rangle  \exp(M^\top(t'- {s}'))
  \label{eq:correlation} \\
  &= \int_0^T d{s} \int_0^{t'} d{s}' \exp(M{s}) \hat R   \exp(M^\top{s}') \, \delta(t-t' -{s} + {s}') \nonumber \\
  &= \int_0^\infty d{s} \exp(M ({s} + (t-t'))) \hat R \exp(M^\top {s} ). \nonumber
\end{align}
We substituted ${s} \rightarrow t - {s}, {s}' \rightarrow t' - {s}'$, used the fact
that
\begin{align*}
  \langle \vec u(t) \vec u^\top(t') \rangle = \begin{pmatrix}
  0 & 0 \\
  0 & Q\langle \vec \xi(t) \vec \xi^\top(t') \rangle Q
\end{pmatrix} = \begin{pmatrix}
0 & 0 \\
0 & QRQ
\end{pmatrix} \delta(t-t') = \hat R\, \delta(t-t'),
\end{align*}
and finally took the limit of $t,t' \rightarrow\infty$ while keeping $t-t'$ fixed.
Since we want to find the variance, we now set $t-t' = 0$.
This matrix-valued integral cannot be evaluated directly, but we can integrate by parts
to obtain
\begin{align}
  \langle \vec y(0) \vec y(0)^\top \rangle = E &= \int_0^\infty d{s} \exp(M {s}) \hat R \exp(M^\top {s}) \nonumber \\
  &= \left[ M^\dagger \exp(M{s})\hat R\exp(M^\top{s}) \right]_0^\infty - M^\dagger\int_0^\infty d{s} \exp(M {s}) \hat R \exp(M^\top {s}) M^\top \nonumber \\
  &= -M^\dagger \hat R - M^\dagger E M^\top \nonumber \\
  \Leftrightarrow ME + EM^\top &= -\hat R.
  \label{eq:lyapunov}
\end{align}
This matrix equation for $E$ is called the \emph{Lyapunov equation}, and there
is no analytic expression for its solution.
(Note that we used the pseudo-inverse. This is allowed because even though $M$ has
a nontrivial nullspace of dimension 1 corresponding to $(\vec{1}, \vec{0})$,
this nullspace is projected out by $\hat R$.)
The total variance of the fluctuations $\vec\varepsilon(t)$ is encoded
in the trace of the upper-left block of $E$.
We write the Lyapunov equation explicitly in block-form,
\begin{align}
  \begin{pmatrix}
    A & B \\
    B^\top & C
  \end{pmatrix}
  \begin{pmatrix}
    0 & -L \\
    \mathbb{1} & -\gamma \mathbb{1}
  \end{pmatrix}
  +
  \begin{pmatrix}
    0 & \mathbb{1} \\
    -L & -\gamma \mathbb{1}
  \end{pmatrix}
  \begin{pmatrix}
    A & B \\
    B^\top & C
  \end{pmatrix}
  =
  \begin{pmatrix}
    0 & 0 \\
    0 & -QRQ
  \end{pmatrix},
  \label{eq:lyapunov-parts}
\end{align}
where we made the symmetric ansatz $E = \begin{pmatrix} A & B \\ B^\top & C \end{pmatrix}$
with $A^\top = A$ and $C^\top = C$
(Remember that $E$ is a correlation matrix and therefore symmetric).
Our goal is now to find an expression for $\operatorname{tr}(A)$.
Multiplying out yields the equations
\begin{align*}
  B + B^\top &= 0\\
  C - AL - \gamma B &= 0 \\
  C - LA - \gamma B^\top &= 0 \\
  2\gamma C + LB + B^\top L &= QRQ.
\end{align*}
Adding and subtracting the first and second yields
\begin{align}
  C &= \frac{1}{2}(AL + LA) \nonumber \\
  B &= \frac{1}{2\gamma}(LA - AL).
  \label{eq:lyapunov-B}
\end{align}
Plugging these into the third and taking the trace,
\begin{align}
  2\gamma \frac{1}{2}(AL + LA) + \frac{1}{2\gamma}(L^2 A - LAL - LAL + A L^2) &= QRQ \nonumber \\
  \Rightarrow \gamma (L^\dagger A L + L^\dagger L A) + \frac{1}{2\gamma}(L^\dagger L^2 A - 2L^\dagger LAL + L^\dagger A L^2) &= L^\dagger QRQ
  \label{eq:lyapunov-A} \\
  \Rightarrow 2\gamma \operatorname{tr}_{\perp\vec{1}}(A) = \operatorname{tr}(L^\dagger QRQ) = \operatorname{tr}(L^\dagger R). \nonumber
\end{align}
Here, we can only take the trace over the subspace perpendicular to the vector with
all ones, because that is the subspace that $L^\dagger$ projects on.
Additionally, we used the fact that $QL = L Q = L$ because
$L$ is a graph Laplacian whose kernel is spanned by
the vector $\vec 1$ of all 1's.

We now show that $\vec{1}^\top A \vec{1} = 0$. We compute directly
\begin{align*}
\vec{1}^\top A \vec{1} &= (\vec 1^\top, 0) E (\vec 1^\top, 0)^\top \\
  &= \int_0^\infty d{s} (\vec 1^\top, 0) \exp(M {s}) \hat R \exp(M^\top {s}) (\vec 1^\top, 0)^\top.
\end{align*}

It is easy to compute
\begin{align*}
  M^\top \begin{pmatrix} \vec 1 \\ 0 \end{pmatrix}
  = \begin{pmatrix} 0 \\ \vec 1 \end{pmatrix} \\
  M^\top \begin{pmatrix} 0 \\ \vec 1 \end{pmatrix}
  = -\gamma \begin{pmatrix} 0 \\ \vec 1 \end{pmatrix}.
\end{align*}
Therefore, the matrix exponential can be expanded into a series,
\begin{align*}
  \exp(M^\top {s}) \begin{pmatrix} \vec 1 \\ 0 \end{pmatrix} &=
  \begin{pmatrix} \vec 1 \\ 0 \end{pmatrix} - \frac{1}{\gamma}
    \begin{pmatrix} 0 \\ \vec 1 \end{pmatrix} \sum_{n=1}^\infty \frac{(-\gamma{s})^n}{n!} \\
      &= \begin{pmatrix} \vec 1 \\ 0 \end{pmatrix} - \frac{1}{\gamma}
        \begin{pmatrix} 0 \\ \vec 1 \end{pmatrix} (e^{-\gamma{s}} - 1).
\end{align*}
From the structure of $\hat R$, we immediately obtain,
\begin{align*}
  \hat R \exp(M^\top {s}) \begin{pmatrix} \vec 1 \\ 0 \end{pmatrix}
  = -\frac{1}{\gamma} \begin{pmatrix} 0 \\ QRQ\vec{1} \end{pmatrix} (e^{-\gamma{s}} - 1) = 0.
\end{align*}

Thus $\vec{1}^\top A \vec{1} = 0$, the trace over the perpendicular subspace is
actually the full trace, and we obtain,
\begin{align}
  \langle |\vec{\varepsilon}(t)|^2 \rangle = \frac{1}{2\gamma} \operatorname{tr}(L^\dagger R).
  \label{eq:total-variance}
\end{align}

\subsection{Colored noise}
We now assume that $\langle \vec \xi(t) \vec \xi(t')^\top \rangle = \frac{\kappa}{2} e^{-\kappa|t-t'|} QRQ$.
(Remember that $\kappa = \tau^{-1}$ is the inverse time scale).
We can express Eq.~\eqref{eq:correlation} as follows, taking the long-time limits,
\begin{align}
  \frac{2}{\kappa}\langle \vec y(0) \vec y(0)^\top \rangle &= \int_0^\infty d{s} \int_0^\infty d{s}'
  e^{-\kappa|{s} - {s}'|} e^{M{s}} \hat R e^{M^\top {s}'} \nonumber \\
  &= \int_0^\infty d{s} \,  e^{(M - \kappa\mathbb{1}){s}} \hat R \int_0^{s} d{s}' e^{(M^\top + \kappa\mathbb{1}) {s}'}
  + \int_0^\infty d{s}\,  e^{(M + \kappa\mathbb{1}){s}} \hat R \int_{s}^\infty d{s}' e^{(M^\top - \kappa\mathbb{1}) {s}'} \nonumber \\
  &= (M-\kappa\mathbb{1})^{-1}\hat R (M^\top+\kappa\mathbb{1})^{-1} + \left(\int_0^\infty d{s}\, e^{M{s}} \hat R e^{M^\top{s}}\right)\left(
  (M^\top+\kappa\mathbb{1})^{-1} - (M^\top-\kappa\mathbb{1})^{-1} \right) \nonumber \\
  &= (M-\kappa\mathbb{1})^{-1}\hat R (M^\top+\kappa\mathbb{1})^{-1} + E \left(
  (M^\top+\kappa\mathbb{1})^{-1} - (M^\top-\kappa\mathbb{1})^{-1} \right),
  \label{eq:ou-correlation-matrix}
\end{align}
where the matrix-valued integral $E$ solves the Lyapunov equation again, Eq.~\eqref{eq:lyapunov}.
In order to continue, we require some expressions for the block-wise inverses
of $M\pm\kappa\mathbb{1}$.
Define $S_{\pm} = (\pm\kappa-\gamma)(\kappa(\kappa\mp\gamma)\mathbb{1} + L)^{-1}$, then
\begin{align*}
  (M \pm\kappa\mathbb{1})^{-1} &= \begin{pmatrix}
  S_\pm & -\frac{S_\pm}{\pm\kappa-\gamma} \\
  \frac{LS_\pm}{\pm\kappa - \gamma} & \frac{\mathbb{1}}{\pm\kappa-\gamma} - \frac{LS_\pm}{(\pm\kappa-\gamma)^2}
\end{pmatrix}\\
(M^\top \pm\kappa\mathbb{1})^{-1} &= (M \pm\kappa\mathbb{1})^{-T}.
\end{align*}
With this, the first term in Eq.~\eqref{eq:ou-correlation-matrix} is
\begin{align}
  (M-\kappa\mathbb{1})^{-1}\hat R (M^\top+\kappa\mathbb{1})^{-1} =
  \begin{pmatrix}
    -\frac{S_{-}QRQS_{+}}{(\kappa+\gamma)(\kappa-\gamma)} & * \\
    * & *
  \end{pmatrix}
  =   \begin{pmatrix}
      (\kappa(\kappa - \gamma)\mathbb{1} + L)^{-1} QRQ (\kappa(\kappa+\gamma)\mathbb{1} + L)^{-1} & * \\
      * & *
    \end{pmatrix},
    \label{eq:ou-first-part}
\end{align}
where we only computed the upper-left block because it contains the correlations
of the fluctuations themselves.
Next, we compute the products of $E$ with the block inverses,
\begin{align}
  E (M^\top \pm \kappa\mathbb{1} )^{-1} = \begin{pmatrix}
  A & B \\
  B^\top & C
\end{pmatrix}
(M \pm\kappa\mathbb{1})^{-1} &=
\begin{pmatrix}
A & B \\
B^\top & C
\end{pmatrix}
\begin{pmatrix}
S_\pm & \frac{LS_\pm}{\pm\kappa - \gamma} \\
 -\frac{S_\pm}{\pm\kappa-\gamma} & \frac{\mathbb{1}}{\pm\kappa-\gamma} - \frac{S_\pm}{(\pm\kappa-\gamma)^2}
\end{pmatrix} \nonumber \\
&=
\begin{pmatrix}
  A S_{\pm} - \frac{B S_\pm}{\pm\kappa-\gamma} & * \\
  * & *
\end{pmatrix},
\label{eq:ou-second-part}
\end{align}
where again we only computed the relevant parts and employed
the decomposition of $E$ from Eq.~\eqref{eq:lyapunov-parts}.
In order to obtain the total fluctuation variance, we need the traces over
the upper-left blocks.
For Eq.~\eqref{eq:ou-first-part}, this is
\begin{align*}
  \operatorname{tr}\left((\kappa(\kappa - \gamma)\mathbb{1} + L)^{-1} QRQ (\kappa(\kappa+\gamma)\mathbb{1} + L)^{-1}\right)
  = \operatorname{tr}\left( ((L + \kappa^2\mathbb{1})^2 - \kappa^2\gamma^2)^{-1} QRQ \right).
\end{align*}
In order to compute the trace in Eq.~\eqref{eq:ou-second-part}, we note that $[S_\pm, L] = 0$
and multiply Eqns.~\eqref{eq:lyapunov-A} and~\eqref{eq:lyapunov-B} by $S_\pm$.
Taking the trace then yields
\begin{align*}
  \operatorname{tr}(A S_\pm) &= \frac{1}{2\gamma} \operatorname{tr}(L^\dagger QRQ S_\pm) \\
  \operatorname{tr}(B S_\pm) &= \frac{1}{2\gamma} \operatorname{tr}((LA-AL)S_\pm) = 0.
\end{align*}
Finally, we obtain
\begin{align*}
  \langle | \vec\varepsilon(t) |^2 \rangle =\frac{\kappa}{2}\left( \operatorname{tr}\left(((L + \kappa^2\mathbb{1})^2 - \kappa^2\gamma^2)^{-1} QRQ \right)
  + \frac{\kappa-\gamma}{2\gamma} \operatorname{tr}(L^\dagger (\kappa(\kappa-\gamma)\mathbb{1} + L)^{-1} QRQ)
  +\frac{\kappa+\gamma}{2\gamma} \operatorname{tr}(L^\dagger (\kappa(\kappa+\gamma)\mathbb{1} + L)^{-1} QRQ) \right).
\end{align*}
This expression can be further simplified by computing the trace in the eigenbasis $\{\phi_i\}$
of $L$,
\begin{align}
  \langle | \vec\varepsilon(t) |^2 \rangle &= \frac{\kappa}{2} \sum_{i>0} \phi^\top_i Q R Q \phi_i
  \frac{\omega_i^2 + \frac{\kappa-\gamma}{2\gamma}(\kappa(\kappa+\gamma) + \omega_i^2)
  + \frac{\kappa+\gamma}{2\gamma}(\kappa(\kappa-\gamma) + \omega_i^2)}{\omega_i^2(\kappa(\kappa-\gamma)+\omega_i^2)(\kappa(\kappa+\gamma)+\omega_i^2)} \nonumber \\
  &= \frac{\kappa(\kappa+\gamma)}{2\gamma} \sum_{i>0} \phi^\top_i R \phi_i \frac{\omega_i^2 + \kappa(\kappa-\gamma)}{\omega_i^2(\kappa(\kappa-\gamma)+\omega_i^2)(\kappa(\kappa+\gamma)+\omega_i^2)}
  \nonumber \\
  &= \frac{\kappa(\kappa+\gamma)}{2\gamma} \operatorname{tr}\left(L^\dagger(L + \kappa(\kappa+\gamma)\mathbb{1})^{-1} R\right) \nonumber \\
  &= \frac{1}{2\gamma} \operatorname{tr}\left(L^\dagger \left[L \frac{1}{\kappa (\kappa + \gamma)}  + \mathbb{1}\right]^{-1} R\right),
  \label{eq:ou-variance}
\end{align}
which corresponds to Eq.~(4) in the main paper using $\kappa=1/\tau$.

\subsection{The Kuramoto model}
The linearized Kuramoto model is described by
\begin{align*}
\dot{\vec{\varepsilon}} + L \vec{\varepsilon} = \vec{\xi}.
\end{align*}
Thus, the calculation from the preceding section
still works upon replacing $M$ by $-L$, and without decomposing into blocks.
The fluctuation variance is simply $\langle |\vec\varepsilon(t)|^2 \rangle
=\operatorname{tr}\left( \langle \vec\varepsilon(t) \vec\varepsilon(t)^\top \rangle \right)$.

In the white noise case we obtain
\begin{align*}
\langle |\vec\varepsilon(t)|^2 \rangle
=\frac{1}{2}\operatorname{tr}\left(L^\dagger R\right),
\end{align*}

and in the case of Ornstein-Uhlenbeck colored noise similarly,
\begin{align*}
\langle |\vec\varepsilon(t)|^2 \rangle
=\frac{\kappa}{2} \operatorname{tr}\left(L^\dagger (L+\kappa\mathbb{1})^{-1} R\right).
\end{align*}

Thus, formally the results for the Kuramoto model and the swing equation are related
by a replacement of variables $\gamma\rightarrow 1, \kappa(\kappa+\gamma) \rightarrow \kappa$
because objective functions that differ only by a constant pre-factor
have the same minimizers.

\section{Numerical Optimization}
Here we describe our optimization algorithm for the case of white noise.
The Ornstein-Uhlenbeck case is similar, with a different objective function.
\subsection{Cost-constrained optimization close to synchrony}
For simplicity, let us consider the case where there are no steady state flows,
$\bar P_{i} = 0$, $\bar\delta_{i} \equiv const$.

We choose to optimize for fixed cost,
\begin{align*}
  N_e C = \sum_e B_e^\alpha,
\end{align*}
where $\alpha$ is a parameter that can be tuned and that controls
the economy of scale for the couplings.
The Lagrangian is
\begin{align*}
  \mathcal{L} = \operatorname{tr}(L^\dagger R) + \lambda \left( \sum_e B_e^\alpha - N_e C\right).
\end{align*}

Taking partial derivatives and setting them to zero yields
\begin{align*}
  \lambda\alpha B_e^{\alpha-1} &= e^\top E^\top L^\dagger R L^\dagger E e \nonumber \\
  \Rightarrow B_e &= c\, (B_e^2 e^\top E^\top L^\dagger R L^\dagger E e)^{\frac{1}{1+\alpha}},
\end{align*}
for a constant $c$.

For $\alpha < 1$, the landscape is non-convex and many local minima exist.
For $\alpha > 1$, the landscape is convex, and one finds a unique global minimum.

\subsection{Cost-constrained optimization with nonzero steady-state flow}
The Lagrangian is again
\begin{align*}
  \mathcal{L} = \operatorname{tr}(L^\dagger R) - \lambda \left( \sum_e B_e^\alpha - N_e C\right).
\end{align*}
The Laplacian weights are $B_e \cos(\Delta \bar{\delta}_{e})$ where the difference $\Delta\bar\delta_e =
\bar\delta_i - \bar\delta_j$ for the edge $e=(ij)$.
We take the derivative with respect to the couplings and set to zero,
\begin{align}
  \lambda\alpha B_e^{\alpha-1} = \cos(\Delta\bar\delta_{e})e^\top E^\top L^\dagger R L^\dagger E e
  - \sum_f B_f \sin(\Delta\delta_{f}) f^\top E^\top L^\dagger R L^\dagger E f \frac{\partial \Delta\bar\delta_{f}}{\partial B_e}.
  \label{eq:fp1}
\end{align}

The derivatives of the steady state angle differences can be computed
by taking derivatives of the steady state condition,
\begin{align*}
   0 &= \frac{\partial}{\partial B_e} \sum_j B_{ij} \sin(\bar\delta_{i} - \bar\delta_{j}) \\
   \Rightarrow \frac{\partial \Delta\bar\delta_{f}}{\partial B_e} &= -f^\top E^\top L^\dagger E e \sin(\Delta\bar\delta_{e}) = -S_{ef} \sin(\Delta\bar\delta_{e}),
\end{align*}
where we defined the symmetric matrix $S_{ef} = e^\top E^\top L^\dagger E f$.
Plugging into Eq.~\eqref{eq:fp1}, we obtain
\begin{align}
  \lambda\alpha B_e^{\alpha-1} = \cos(\Delta\bar\delta_{e})\langle \Delta\varepsilon^2_e \rangle
  + \sin(\Delta\bar\delta_{e}) \sum_f S_{ef} B_f \langle \Delta\varepsilon^2_f \rangle_R \sin(\Delta\bar\delta_{f}).
  \label{eq:fp2}
\end{align}
Here, we introduced the shorthand $\langle \Delta\varepsilon^2_e \rangle_R = e^\top E^\top L^\dagger R L^\dagger E e$
for the average squared linearized angle difference along an edge under the correlation matrix $R$.

For Ornstein-Uhlenbeck correlations, a similar but more unwieldy expression holds.

\subsection{Algorithm for cost-constrained optimization}
In order to solve Eq.~\eqref{eq:fp2}, we use the following algorithm, based on Ref.~[39].
\begin{enumerate}
  \item Start with initial couplings $B_e^{(0)}$
  \item Run a few steps of a nonlinear root finder (trust-region algorithm as implemented in the package \texttt{NLsolve.jl}, \texttt{https://github.com/JuliaNLSolvers/NLsolve.jl}.)
  to obtain the steady state angles $\delta_{ss}^{(0)}$
  \item Compute
  \begin{align*}
    \hat B_e^{(i+1)} = \left((B_e^{(i)})^2\cos(\Delta\bar\delta_{e}^{(i)})\langle \Delta(\varepsilon^2_e)^{(i)} \rangle
    + (B_e^{(i)})^2\sin(\Delta\bar\delta_{e}^{(i)}) \sum_f S_{ef}^{(i)} B_f^{(i)} \langle (\Delta\varepsilon^2_f)^{(i)} \rangle_R \sin(\Delta\bar\delta_{f}^{(i)})
    \right)^{\frac{1}{1+\alpha}}
  \end{align*}
  \item Normalize
  \begin{align*}
    B_e^{(i+1)} = C^{1/\alpha}\frac{\hat B_e^{(i+1)}}{\left(\sum_f (\hat B_f^{(i+1)})^\alpha\right)^{1/\alpha}}
  \end{align*}
  \item Run another few iterations of a nonlinear root finder to obtain $\bar\delta^{(i+1)}$
  \item Repeat from 3 until convergence of both the steady state angles
  and the $B_e$.
\end{enumerate}
Sometimes the RHS in step 3 becomes negative in an intermediate step.
In that case we set it to zero hoping to converge to a good solution later.

The number of variables in the cost-constrained optimization
is given by the number of nodes $N$ in the network, for which the non-linear root finder in step 2 and 5 solves, and by the number of edges $E$ which are obtained by the fixed-point iteration in steps 3 and 4.
In regular graphs such as the ones we consider, each node is connected
by the same number of edges (except at the boundaries), such that $E = \mathcal{O}(N)$.
For a regular network constructed from rows containing $n$ nodes each in $d$ spatial dimensions,
the number of variables thus scales as $\mathcal{O}(n^d)$.

\section{Improvement due to optimization}
\begin{figure}[h]
  \centering
  \includegraphics[width=.3\textwidth]{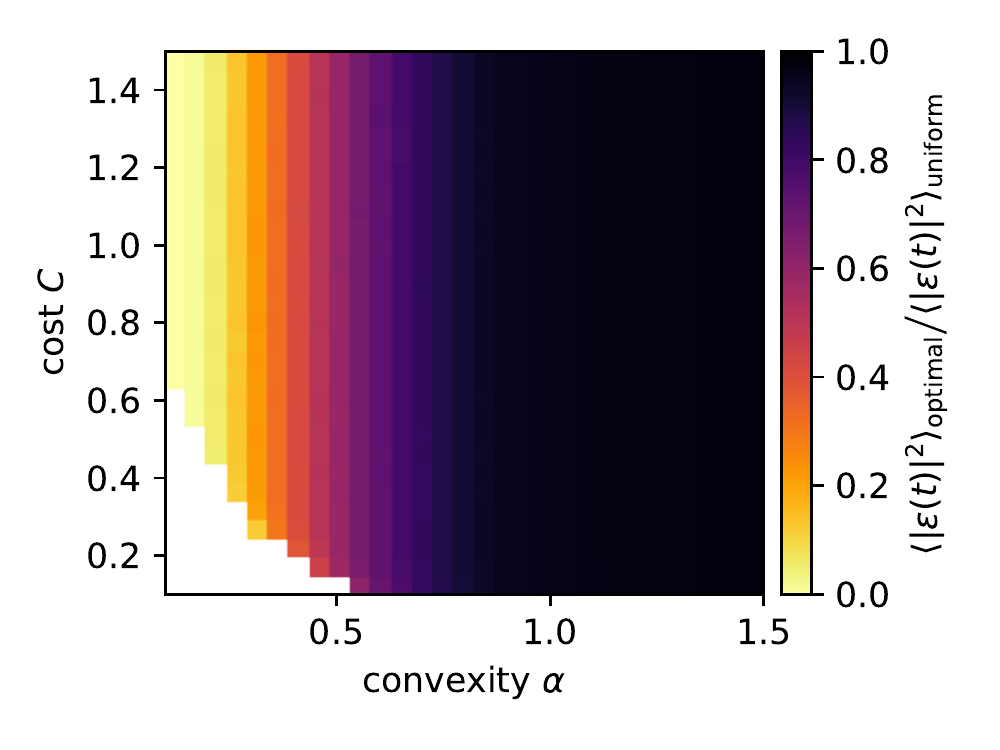}
  \caption{Improvement in noise canceling due to optimization.
  We compute the ratio of the optimal network objective
  $\langle |\vec\varepsilon(t)|^2 \rangle_{\mathrm{optimal}}$
  to the objective $\langle |\vec\varepsilon(t)|^2 \rangle_{\mathrm{uniform}}$, which is computed for
  uniform networks, $B_{ij} = const$.
  Each pixel is an average over ratios computed for 25
  optimal networks with different, uniformly random
  initial conditions.
  While the improvement is insignificant for convex,
  dense networks (which are almost uniform anyways),
  sparse networks with $0<\alpha<1$
  provide a significant advantage.}
  \label{fg:improvement}
\end{figure}

\section{Time series of the non-linear swing equation dynamics}
We solve the non-linear swing equation with stochastic feed-in
as a system of coupled stochastic differential equations.
In SDE form the white noise case reads,
\begin{align*}
d\delta_i &= \nu_i dt \\
d\nu_i &= -\gamma \nu_i dt + \sum_{j=1}^N B_{ij} \sin(\delta_i - \delta_j)dt + \bar P_i^c dt +
\sum_{j=1}^N C_{ij} dW_j,
\end{align*}
where the $dW_j$ are i.i.d.~Wiener processes and $C = U \sqrt{\Sigma}$
is constructed from the singular value decomposition of the correlation
matrix, $R = U \Sigma U^\top$. With this definition, the feed-ins have the
desired correlation matrix
$\langle C \frac{d\vec{W}(t)}{dt} \frac{d\vec{W}(t')^\top}{dt} C^\top \rangle= R \,\delta(t-t')$.

In the Ornstein-Uhlenbeck case the system of SDEs is augmented to
\begin{align}
dX_i &= -\kappa X_i dt + \sqrt{2\kappa}dW_i \label{eq:ou-eqn}\\
d\delta_i &= \nu_i dt \nonumber \\
d\nu_i &= -\gamma \nu_i dt + \sum_{j=1}^N B_{ij} \sin(\delta_i - \delta_j)dt + \bar P_i^c dt +
\sum_{j=1}^N C_{ij} X_j dt, \nonumber
\end{align}
where again the $dW_j$ are i.i.d. Wiener processes and the matrix $C$
is defined as before. We obtain the desired feed-in correlations,
$\langle C \vec{X}(t) \vec{X}(t')^\top C^\top \rangle= R\, e^{-\kappa |t-t'|}$.
We employ the Julia language's \texttt{DifferentialEquations.jl} package
to solve the SDEs using the Euler-Maruyama method.
Eq.~\eqref{eq:ou-eqn} is replaced by the package's time step-independent
distributionally correct Ornstein-Uhlenbeck process.

We then define the instantaneous mean angle by the integral
\begin{align*}
\bar \delta_i(t) = \frac{1}{t}\int_0^t \delta_i(t') dt',
\end{align*}
where we evaluate the integral numerically from the simulation time series.
From this, the numerical fluctuations and fluctuation variances are
\begin{align*}
\varepsilon_i(t) &= \delta_i(t) - \bar \delta_i(t) \\
\langle |\vec\varepsilon(t)|^2 \rangle &= \frac{1}{t}\int_0^t \left|\vec\delta(t) - \bar{\vec\delta}(t)\right|^2 dt'.
\end{align*}

\begin{figure}[h]
  \centering
  \includegraphics[width=.8\textwidth]{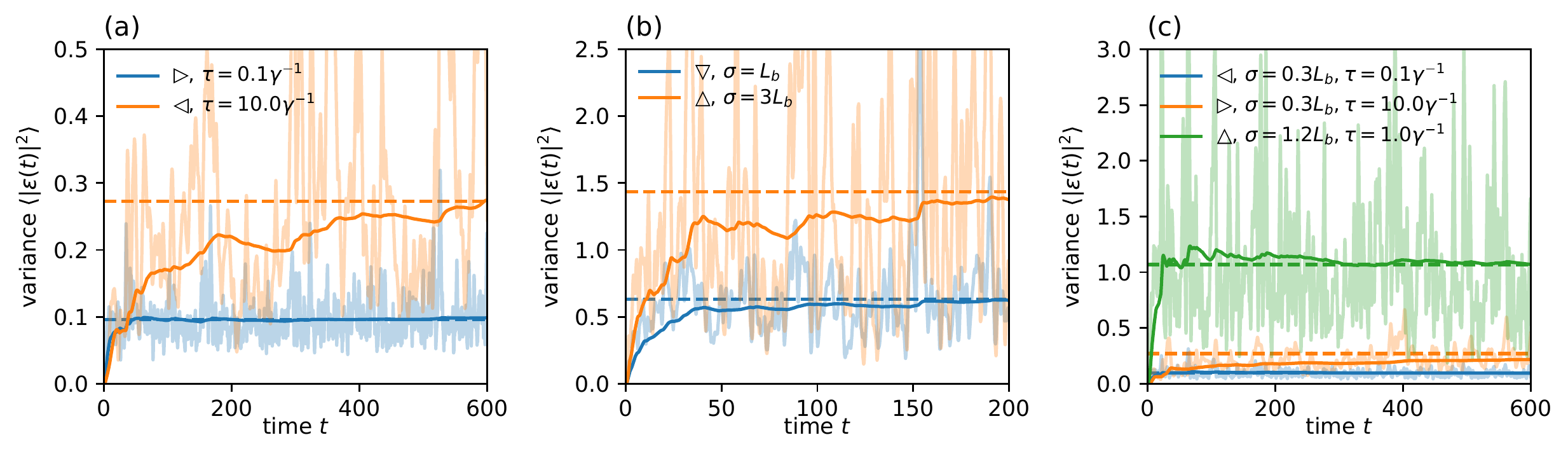}
  \caption{Time series of the non-linear swing equation dynamics
  for the same networks as in the main paper, Fig.~2.
  (a) Ornstein-Uhlenbeck colored noise and spatially incoherent feed-in.
  (b) Gaussian spatially coherent feed-in with temporal white noise.
  (c) Spatio-temporally correlated feed-in.}
  \label{fg:time-series}
\end{figure}
In addition to the time series for white noise and spatially
incoherent noise shown in the main paper, Fig.~1,
here we also show time series for the optimal networks
from Figs.~2 and 3 of the main paper (see Fig.~\ref{fg:time-series}).
It can be seen clearly that higher correlations also lead
to larger fluctuation variances.

\subsection{Validity of the linear model}
\begin{figure}[h]
  \centering
  \includegraphics[width=.6\textwidth]{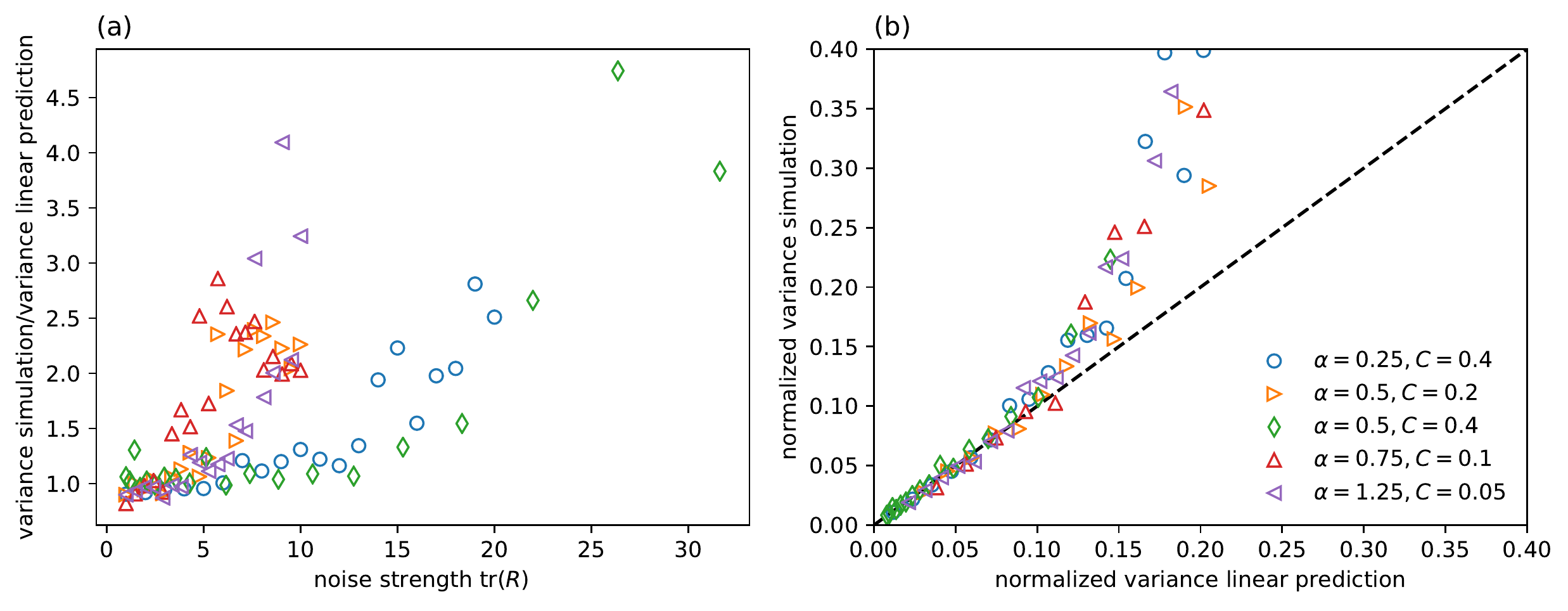}
  \caption{Validity of the linear optimized model for white noise.
  (a) We compare optimal networks at various values
  of convexity $\alpha$ and cost $C$ by plotting the ratio
  of the total fluctuation variance computed as a time average
  from fully non-linear simulations run until a time $t=200$
  and the prediction from the linear
  model. For each $8\times 8$ triangular network,
  we vary the total noise strength $\operatorname{tr}(R)$, where
  $R$ is spatially incoherent. The linear approximation
  is good for small noise strengths.
  (b) Normalizing the variance by the worst case variance,
  $N \pi^2/3$, computed by assuming uniformly distributed fluctuations
  $\varepsilon_i$ on $[-\pi,\pi)$.
  The linear prediction is adequate until the predicted variance reaches
  approximately $10\%$ of the worst case variance.}
  \label{fg:validity}
\end{figure}

In order to test the validity of the linear model,
we perform simulations of the fully nonlinear swing equation
in the white noise, spatially incoherent case for various values
of $C$ and $\alpha$. For each combination of parameters,
we scale the total noise variance $\operatorname{tr}(\hat R)$
until the linear prediction and simulations begin to disagree
(see Fig.~\ref{fg:validity}~(a)).
By rescaling the total noise variances
$\langle |\vec\varepsilon(t)|^2\rangle \rightarrow \langle |\vec\varepsilon(t)|^2\rangle/(N\pi^2/3)$, where
$N\pi^2/3$ is the worst case variance, we see that the linear
model is accurate up to $\approx 10\%$ of the worst case variance
(see Fig.~\ref{fg:validity}~(b)).

\section{Dependence of Ornstein-Uhlenbeck topology on cost}
\begin{figure}[h]
  \centering
  \includegraphics[width=.5\textwidth]{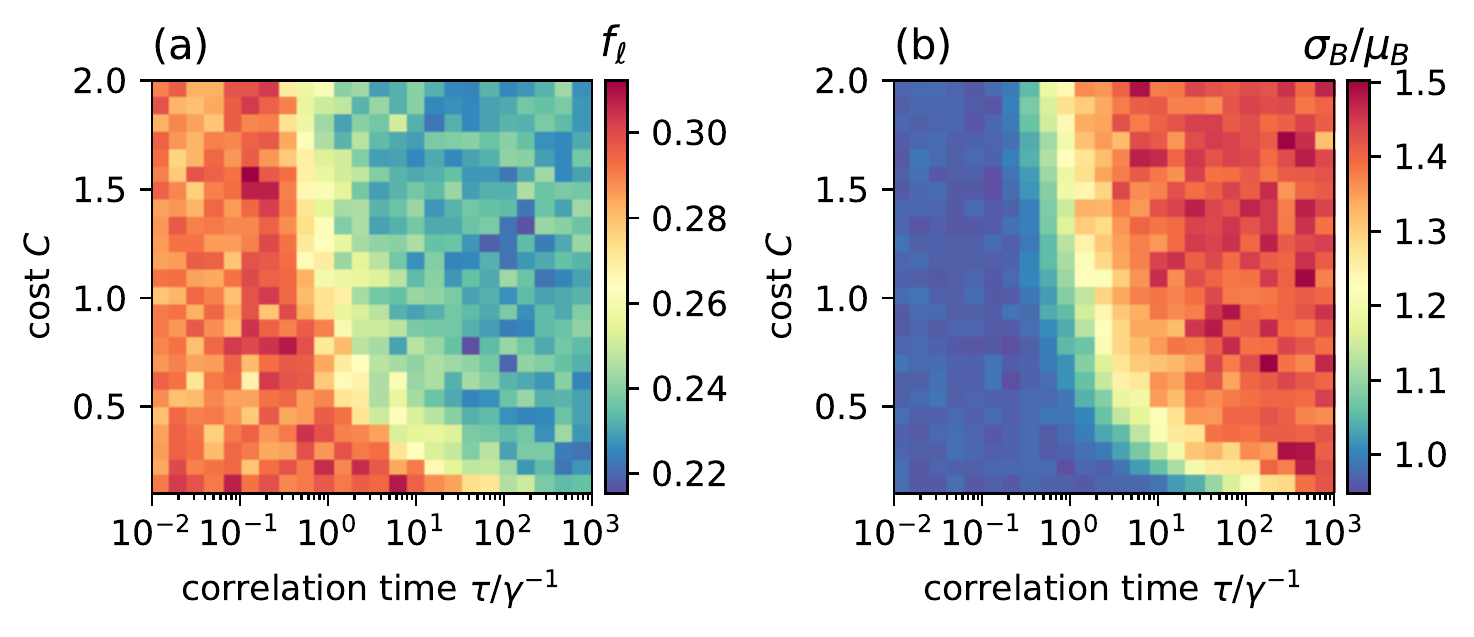}
  \caption{Dependence of the topology of Ornstein-Uhlenbeck optimized networks
  on the cost parameter. We show phase space of the loop density
  and coupling spread for
  $8\times 8$ triangular networks at fixed $\alpha=0.5$ and for spatially
  incoherent noise.
  Very low cost networks with $C\ll 1$ (which have small couplings) stay uniform
  and dense even at longer correlation times $\tau$.
  For higher cost networks with $C > 1$ with larger couplings the transition to
  sparsity and hierarchical organization occurs for smaller $\tau$.}
  \label{fg:ou-cost}
\end{figure}
Unlike for the white noise case, the Ornstein-Uhlenbeck noise variance
Eq.~\eqref{eq:ou-variance} is not homogeneous upon rescaling the
cost parameter $C\rightarrow sC$, even in the well-synchronized limit
$\bar\delta_i \approx 0$. Therefore, unlike for white noise, the optimal
networks depend on $C$.
Fig.~\ref{fg:ou-cost} shows the phase space of
optimal networks as a function of cost and correlation time.
We see that while for small $C$ the transition between topologies
shifts towards larger ${s}$, the topologies themselves remain
unchanged (as quantified by $f_\ell$ and $\sigma_B/\mu_B$).

\section{Phase spaces and optimal networks for square grids}
\begin{figure}[h]
  \centering
  \includegraphics[width=.4\textwidth]{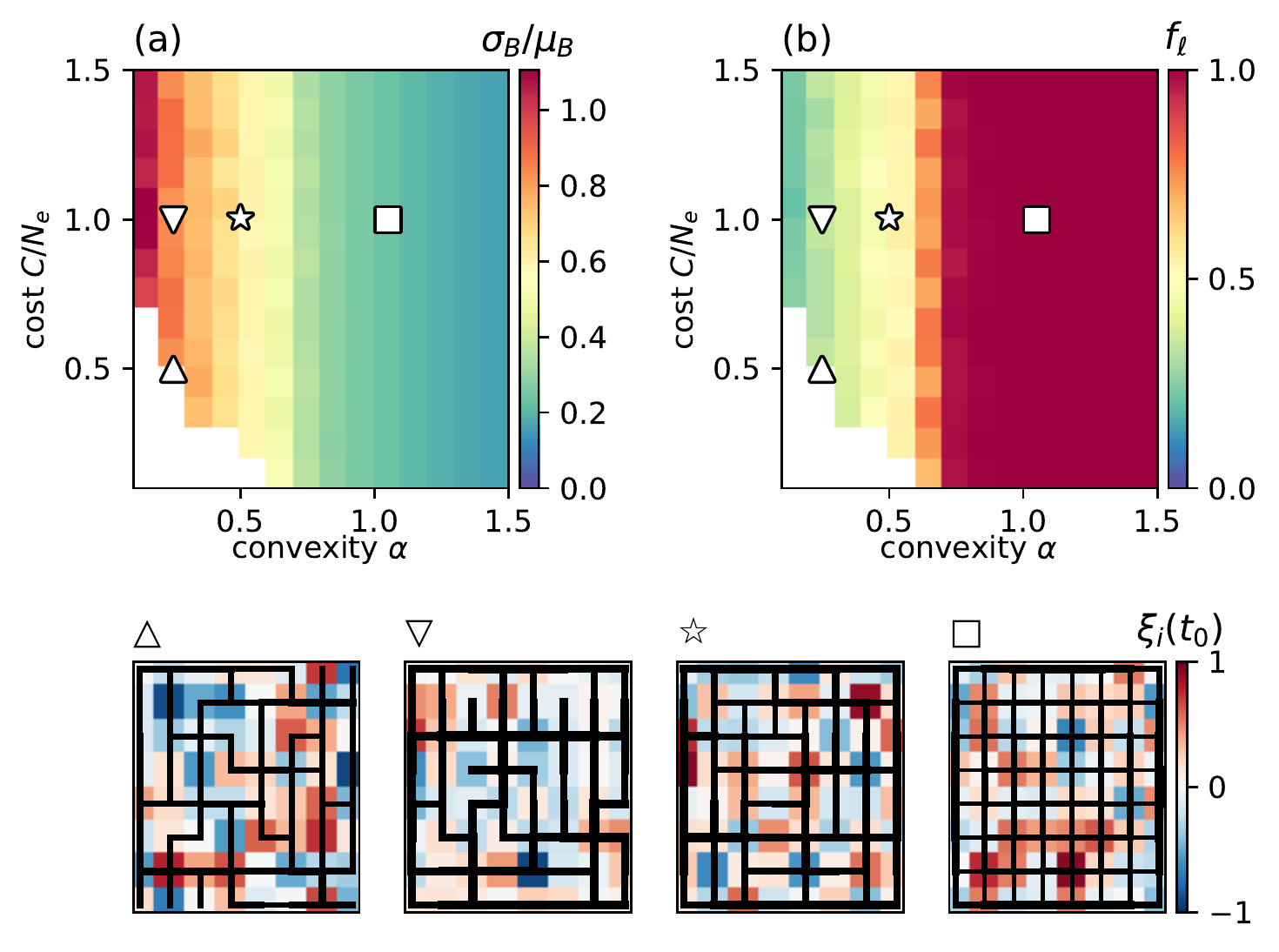}
  \caption{Topology phase space and optimal networks for $8\times 8$ square grids for white noise in time
  and incoherent spatial feed-in.
  Each pixel in the $15\times 15$ phase space is an average over
  5 optimal networks.}
  \label{fg:wn-square}
\end{figure}

\begin{figure}[h]
  \centering
  \includegraphics[width=.6\textwidth]{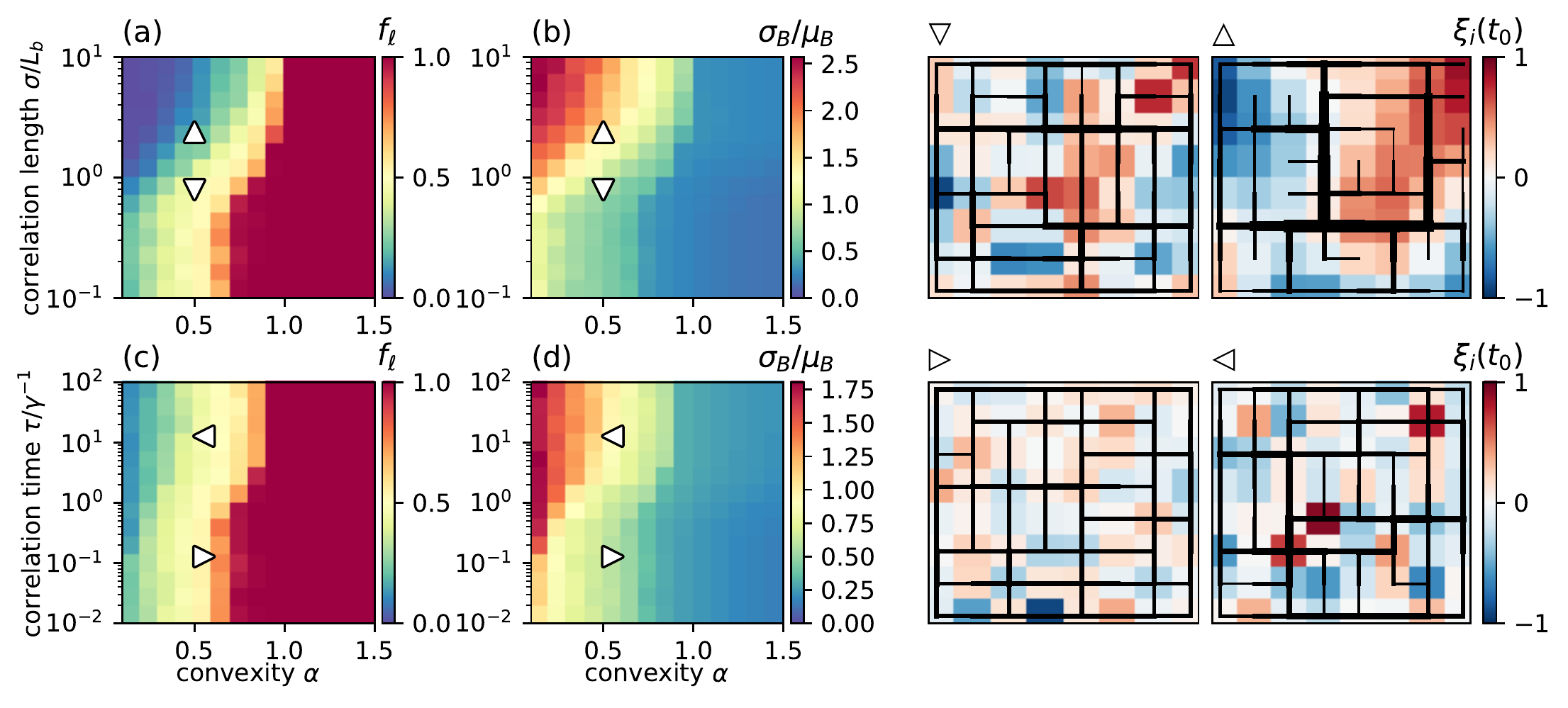}
  \caption{Topology phase space and optimal networks for $8\times 8$ square grids in the well-synchronized limit
  at $C=1$. (a,b) White noise in time
  and Gaussian correlated spatial feed-in.
  (c,d) Ornstein-Uhlenbeck noise in time and
  spatially incoherent feed-in.
  Each pixel in the $15\times 15$ phase space is an average over
  5 optimal networks.}
  \label{fg:corr-square}
\end{figure}

\begin{figure}[h]
  \centering
  \includegraphics[width=.4\textwidth]{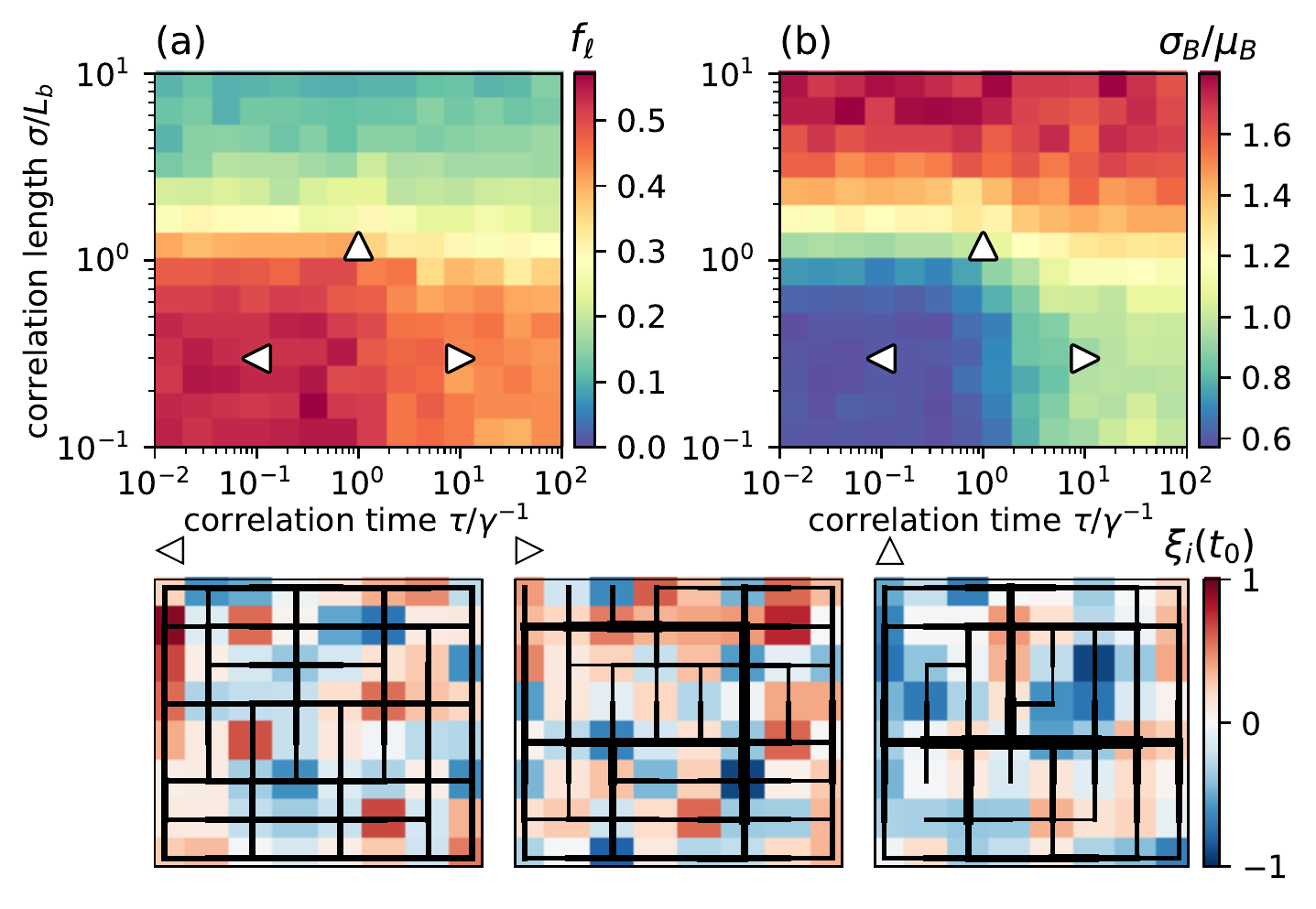}
  \caption{Topology phase space and optimal networks for $8\times 8$ square grids with spatio-temporal correlated
  feed-in in the well-synchronized limit at $C=1$, $\alpha=0.5$.
  Each pixel in the $15\times 15$ phase space is an average over
  5 optimal networks.}
  \label{fg:spatiotemporal-square}
\end{figure}

\end{document}